\documentclass[aps,prl,twocolumn,groupedaddress,amsfonts,amsmath,boldmath]{revtex4}
\usepackage{graphicx}

\usepackage[applemac]{inputenc}
\usepackage[T1]{fontenc}

\newcommand{\rr}{{\bf r}}
\newcommand{\pp}{ {\bf p}}

\newcommand{\uu}{{\bf u}}

\newcommand{\kk}{{\bf k}}
\newcommand{\RR}{{\bf R}}
\newcommand{\PP}{{\bf P}}
\newcommand{\QQ}{{\bf Q}}

\begin{document}
\title{Hydrodynamics of confined active fluids}
\author{Tommaso Brotto$^1$}
\author{Jean-Baptiste Caussin$^1$}
\author{Eric Lauga$^2$}
\author{Denis Bartolo$^{1,3}$}
\affiliation{$^1$PMMH ESPCI-ParisTech-CNRS UMR 7636-Universit\'e Pierre et Marie Curie-Universit\'e Paris Diderot,10 rue Vauquelin 75231 Paris cedex 05, France.}
\affiliation{$^2$Department of Mechanical and Aerospace Engineering, 
University of California San Diego, 9500 Gilman Drive,  La Jolla, CA 92093-0411, USA.}
\affiliation{$^3$Universit\'e de Lyon, Laboratoire de Physique, Ecole Normale Sup\'erieure de Lyon, CNRS UMR 5672, 46 All\'ee d'Italie, F-69364 Lyon cedex 07, France.}
\begin{abstract}
We theoretically describe the dynamics of swimmer populations confined in thin liquid films. We first demonstrate that  hydrodynamic interactions  between confined  swimmers  only depend on their shape and are independent of their specific swimming mechanism. We also show that due to friction with the walls, confined swimmers do not reorient due to flow gradients but the flow field itself. We then quantify  the consequences of these microscopic interaction rules on the large-scale  hydrodynamics of  isotropic populations.  We investigate in details their stability and the resulting phase behavior, highlighting the differences with conventional active, three-dimensional suspensions. Two classes of polar swimmers are distinguished depending on their geometrical polarity.  The first class gives rise to coherent directed motion at all scales  whereas for the second class  we predict the spontaneous formation of coherent clusters (swarms).
\end{abstract}
\maketitle


Soft materials composed of motile particles have seen a surge of interest over the last couple of years. They encompass  auto-phoretic colloids~\cite{Theurkauff:2012jo}, self-propelled droplets~\cite{Thutupalli:2011bv}, and vibrated grains~\cite{Kudrolli:2010hg,Deseigne:2010gc}. This interest was triggered by their  fascinating  structural and transport properties akin to the one found in biological systems such as  bacterial suspensions, migrating cells, and cytoskeletal extracts (see Ref.~\cite{Marchetti:2012ws} and references therein). These so-called active fluids are ensembles of self-driven particles capable of  propelling themselves in the absence of any  external actuation~\cite{Toner:2005wy,Ramaswamy:2010bf,Marchetti:2012ws,Vicsek:2012gp,Cates:2012is}.  From a theoretical perspective, these systems are commonly separated into two classes depending on the way they exchange momentum with their surroundings~\cite{Toner:2005wy,Ramaswamy:2010bf,Marchetti:2012ws}. "Dry" systems, typically walkers, or crawlers, achieve locomotion by transferring momentum to a rigid substrate, and interact via short range contact interactions. In contrast "wet" systems, typically suspensions of swimmers, conserve momentum, and the particles interact at finite distance via long-range hydrodynamic interactions. A number of experimentally relevant situations  involve monolayers of active particles living in confined fluid films, and thus belong to both classes --  e.g.  bacteria swimming on the surface of a cell-culture gel, or active colloids and droplets moving in  microfluidic channels~\cite{Thutupalli:2011bv,Darnton:2010dn,Zhang:2010jn}. 

 In this letter, we describe the phase behavior of  active fluids confined in two-dimensional (2D) geometries. In order to do so, we first revisit the description of  hydrodynamic interactions under confinement. We demonstrate that the far-field flow induced by a swimmer does not depend on the specifics of its swimming mechanism. The notions of pushers and pullers for instance, prevalent in three dimensions (3D), are not relevant in thin films~\cite{Saintillan:2007jr,Saintillan:2008jt}. In addition, on the basis of a prototypal microscopic model, we show that due to friction with the walls, confined polar swimmers are not only prone to align along the local elongation axis but with the flow field itself. We then exploit these new interactions rules in 2D to address   the large-scale dynamics of  confined populations of swimmers. We establish a novel set of hydrodynamic equations for confined active films, which qualitatively differ from the modified Leslie-Eriksen equations for active liquid crystals~\cite{Marchetti:2012ws}. 
An investigation of the resulting phase behavior   leads to the distinction between two classes of polar swimmers depending on their geometrical polarity. The first class (large-head), gives rise to the emergence of coherent particle motion along the same direction at all scales whereas for the  second class  (large-tail), we predict the spontaneous formation of coherent clusters (swarms). 
 
\begin{figure}[b]
\begin{center}
 \includegraphics[width=\columnwidth]{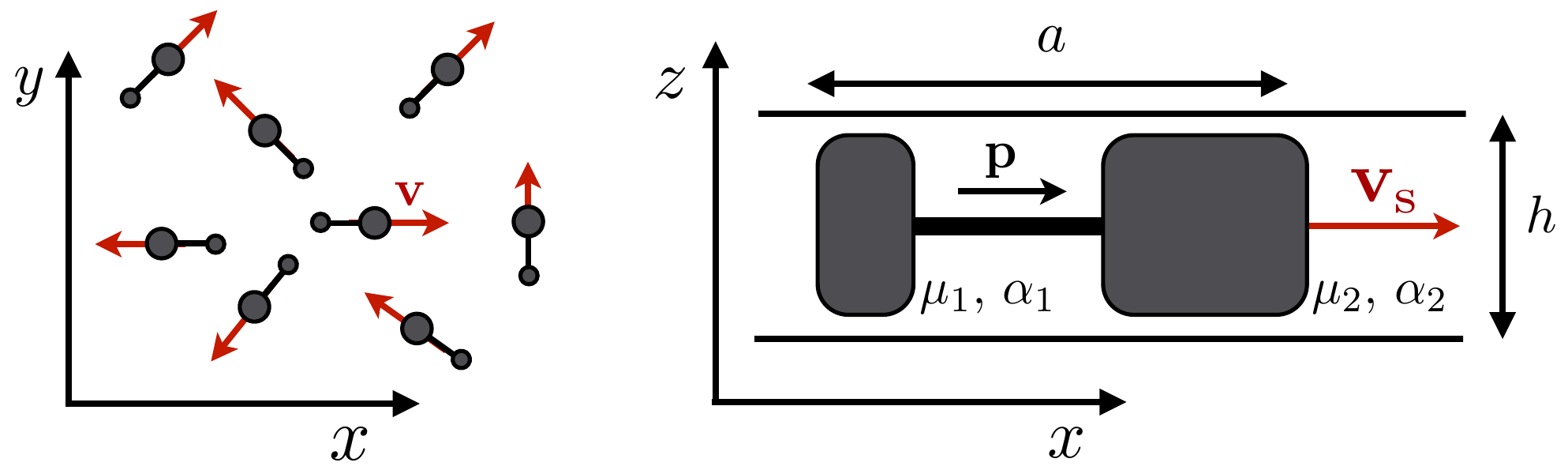}
\caption{Left: Sketch of a confined suspension of active particles swimming freely  in the $(x,y)$ plane. Right: Close-up on a single polar swimmer (see text for notation). The active particles are confined between two walls in the $z$-direction.}
\label{fig1}
\end{center}
\end{figure}
Let us consider an ensemble of self-propelled particles confined in a thin film of a Newtonian liquid. 
We address  strongly confined geometries where the particle height is comparable to the film thickness, $h$, see Fig.~\ref{fig1} (left). 
At scales larger than $h$, the fluid flow is characterized by  the projection of the $z$-averaged velocity field in the $(x,y)$ plane. Far from a swimmer, the projected flow field $\uu(\rr,t)$ is potential
\begin{equation}
\uu(\rr)=-G\nabla \Pi(\rr),
\label{potential}
\end{equation}
where  $\Pi(\rr)$ is the pressure at $\rr=(x,y)$. The Darcy factor $G$  scales as $G\sim h^2/\eta$~\cite{Hulin}. 

How does confinement  affect hydrodynamic interactions between swimmers? 
In unbounded fluids, the flow induced by a swimmer depends on the microscopic details of the propulsion mechanism~\cite{Lauga:2009ul,Drescher:2010cz,Drescher:2011gd}. In the far-field, this flow is often well approximated by a force-dipole singularity, with a $\sim1/r^2$ spatial decay, and as such has been used in most  theoretical models~\cite{Saintillan:2008jt,Baskaran:2009vu,Leoni:2010bh}. This description results in the distinction between so-called pushers (or extensile swimmers), and pullers (or contractile swimmers). They corresponds to {\em force}-dipoles having opposite signs, and displaying  different large scale dynamics~\cite{Saintillan:2008jt,Baskaran:2009vu,Leoni:2010bh}. When confined by solid walls, these  flows are screened algebraically and decay as $\sim1/r^3$, while retaining their  angular symmetry. This screening of hydrodynamic interactions  was shown to  suppress generic instabilities which are the hallmark of isotropic pusher suspensions~\cite{Marchetti:2012ws}. 

As it turns out, however, the two main consequences of confinement has actually been overlooked so far. Any multipolar stress distribution on the surface of the swimmer actually yields only subdominant contributions to the flow in the far field. For any  particle transport mechanism (swimming, driving, advection) the far-field flow induced by a  particle moving in a confined fluid has instead the symmetry of a potential {\em source}-dipole and decays as $\sim1/r^2$~\cite{Beatus:2006hc,Champagne:2010dm,Liron:1976wz}.  The distinction between pushers and pullers is thus irrelevant under confinement.  Irrespective of the propulsion mechanism,  the flow induced by  a swimmer located at $\rr=\RR(t)$ is defined by Eq.~\eqref{potential} and by a modified incompressibility relation
\begin{equation}
\nabla\cdot \uu(\rr)=-\boldsymbol \sigma\cdot\nabla\delta(\rr-\RR(t)),
\label{incompressibility}
\end{equation}
where the dipole strength is  $\boldsymbol \sigma\equiv \sigma\left[\dot\RR(t)-\uu^{(0)}(\RR(t))\right]$ where $\uu^{(0)}$ is the  velocity field in absence of the particle, and  $\sigma$ scales as the square of the particle size (for a disk-shape particle, $\sigma$ is twice the particle area)~\cite{Beatus:2006hc}. 
 The dipolar solution, $\uu^{\rm d}(\rr|\RR(t), \boldsymbol \sigma)$, of Eqs.~\eqref{potential}-\eqref{incompressibility} is given, for a particle located at the origin, by
  \begin{equation}
\uu^{\rm d}(\rr|{\bf 0},\boldsymbol \sigma)=
\frac{1}{2\pi |{\rr}|^2}\left(2{\hat \rr \hat \rr} - {\bf I}\right) \cdot \boldsymbol \sigma,
\label{singledipole} 
\end{equation}
 with $\hat \rr\equiv\rr/|\rr|$ and ${\bf I}$ the identity tensor~\cite{Beatus:2006hc,Liron:1976wz}.  This framework  has proven to accurately describe the interactions between confined advected droplets even in concentrated systems~\cite{Beatus:2009dg,Beatus:2007ju,Champagne:2010dm,Champagne:2011fz}. Importantly, the angular symmetry of $\uu^{\rm d}$ is  different from the one of a force dipole: it is a polar flow field displaying  the same angular dependence  as that of a force {\em monopole} under confinement~\cite{Liron:1976wz} despite the swimmers being self-driven. The reason for this apparent paradox lies in the continuous  momentum exchange with the  confining walls, via the shear flow in the thin films that lubricate the swimmer-wall contacts, see Fig.~\ref{fig1}. 
 
The second important difference with 3D suspensions concerns hydrodynamic interactions between swimmers. In order to  account for these  interactions,  we first establish the equations of motion of an isolated swimmer in a arbitrary fluid flow.  We focus on swimming bodies with  polar shapes, as is the case for most  motile cells.  For a swimmer at position $\RR(t)$ we denote  $\pp(t)$ its orientation ($|\pp|^2=1$) and  $v_{\rm s}$ the  magnitude of its swimming velocity along $ \pp$. From symmetry considerations and at leading order in $|\nabla \uu |$, 
the equations of motion of a polar swimmer for $\{\RR(t),\pp(t)\}$ take the generic form
\begin{eqnarray}
\dot R_\alpha &= &v_{\rm s} p_\alpha + \mu_\perp (\delta_{\alpha\beta}-p_\alpha p_\beta ) u_\beta + \mu_\parallel (p_\alpha p_\beta) u_\beta, 
 \label{eqposition}\\
\dot p_\alpha&= &\nu (\delta_{\alpha \beta} - p_\alpha p_\beta) u_\beta+\nu'(\delta_{\alpha \beta} - p_\alpha p_\beta)(\nabla_\gamma u_\beta) p_\gamma,\quad
\label{eqorientation}
\end{eqnarray}
where $\mu_\perp$ (resp.~$\mu_\parallel$) is a transverse (resp.~longitudinal) mobility coefficient and  $\nu$ and $\nu'$ are two rotational mobility coefficients. In unbounded fluids, we have $\nu=0$ and $\mu_\perp=\mu_\parallel=1$, and  Eq.~\eqref{eqorientation} then corresponds to Jeffrey's equation commonly used to quantify the orientation of anisotropic particles with the flow-elongation axis~\cite{Lauga:2009ul,Saintillan:2008jt}. Conversely, confined suspensions offer the  possibility of having a nonzero value for $\nu$.  Instead of reorienting due to flow gradients, swimmers can reorient because of the flow itself, a new type of  orientational dynamics which has not been considered so far. 

To provide insight into the conditions for nonzero values of $\nu$, we derive   the above equations of motion  for a prototypal  microscopic model (dumbbell). We show how the lubricated friction with the walls induce both  anisotropic mobility ($\mu_\perp\neq \mu_\parallel$) and a direct coupling between the flow velocity and the particle orientation ($\nu\neq0$). Consider a rigid-dumbbell swimmer, composed of two disks of radius $b_1$ (resp.~$b_2$) located at $\RR_1$ (resp.~$\RR_2$), and connected by a frictionless rigid rod of length $a\gg \{b_1,\,b_2\}$ (see Fig.~\ref{fig1}, right).  The lubrication forces between a disk-shape particle and the solid walls  hinder its advection by the fluid. Passive disks would be transported at a velocity $\dot\RR_i(t)=\mu_i \uu(\RR_i)$ ($i=1,2$), where  the  mobility coefficient $\mu_i$ is comprised between 0 (fixed obstacle) and 1 (passive tracer). We also introduce the drag coefficients $\alpha_i$: when a disk is  pulled by an external force $\bf F$ in a quiescent fluid, it moves at a velocity $\dot\RR_i(t)=\alpha_i \bf F$. Let us now assume that the two disks would propel at a velocity $v_{\rm s}^{(0)}\pp$ when alone, and let us compute the swimming speed and mobility coefficients from Eqs.~\eqref{eqposition}-\eqref{eqorientation} for the dummbell. The displacement of each disk results from the competition between (i) self-propulsion, (ii) the advection by the external flow $\uu^{(0)}$, (iii)  the advection of the disk $i$ by the dipolar perturbation induced by the motion of the  disk $j$, $\uu^{\rm d}(\RR_i|\RR_j,\boldsymbol \sigma_j)$, and (iv) the inextensibility constraint, $\RR_2-\RR_1=a\pp$. At leading order in $b_i/a$, these contributions yield the   following equations of motion for the "head" ($i=2$) and the "tail" ($i=1$) of the swimmer:
\begin{eqnarray}
\dot \RR_{1} &=& v_{\rm s}^{(0)}\pp+\mu_1 [\uu^{(0)}({\RR_1}) + \uu^{\rm d}({\RR_1|\RR_2,\boldsymbol \sigma_2})] + \alpha_1 {\bf T} ,
\label{r1}\\
\dot \RR_{2} &=& v_{\rm s}^{(0)}\pp+\mu_2 [\uu^{(0)}({\RR_2}) + \uu^{\rm d}({\RR_2|\RR_1,\boldsymbol \sigma_1})] - \alpha_2 {\bf T} , \quad \,\,
 \label{r2}
\end{eqnarray}
where the tension $\bf T$ ensures the inextensibility condition, $\pp\cdot(\dot \RR_2-\dot \RR_1)=0$. Defining the center of drag of the swimmer as $\RR\equiv (\alpha_1\RR_2+\alpha_2\RR_1)/(\alpha_1+\alpha_2)$, Eqs.~\eqref{r1}-\eqref{r2} are readily recast into the form of Eqs.~\eqref{eqposition}-\eqref{eqorientation} with a  dumbbell velocity and  mobility coefficients  given at leading order by 
 $v_{\rm s}=v_{\rm s}^{(0)}+{\cal O}((b_i/a)^2)$,
$\mu_\perp = \alpha_2 \mu_1 (1-\gamma_2) +\alpha_1 \mu_2 (1-\gamma_1)$, $\mu_\parallel = \alpha_2 \mu_1 (1+\gamma_2) + \alpha_1 \mu_2 (1+\gamma_1)$ and $\nu = [(\mu_2+\mu_1\gamma_2)-(\mu_1+\mu_2\gamma_1)]/a$, where  $\gamma_i\equiv b_i^2(\mu_i-1)/a^2$. 
We first see   that the translational mobility coefficients, $\mu_{\perp,\parallel}$ depend only on the anisotropy of the swimmer, and are independent of its geometrical polarity (they remain unchanged upon a $1 \leftrightarrow 2$ permutation). In addition, as $\mu_\parallel<\mu_\perp$,  a non-swimming dumbbell making a finite angle with a uniform flow field would drift at a finite angle from the flow direction. We also obtain that indeed $\nu\neq 0$   for polar swimmers. Since the   $\mu_i$'s are decreasing functions of the particle radius,  $\nu$ is negative for large-head swimmers ($b_2>b_1$), and positive otherwise. From Eq.~\eqref{eqorientation} we thus get that in a uniform flow  large-head swimmers would reorient against the flow and propel upstream. In contrast,  large-tail swimmers ($b_1>b_2$) would swim downstream. For apolar swimmers, $\nu$ vanishes and the orientation of a symmetric dumbbell evolves according to the Jeffrey's orbits, Eq.~\eqref{eqorientation}, where $\nu=0$ and $\nu'=a [\mu_2(1+\gamma_1) + \mu_1(1+\gamma_2)]/2$.  Note that since $\uu$ is irrotational, the orientation of an isotropic swimmer made of a single disk is not coupled to the background flow. 
In the rest of the paper we discard the conventional $\nu'$ contribution to the orientational dynamics.  It  only yields short-wavelength corrections to the large-scale description of polar-swimmers suspensions described below.

We now turn to the dynamics of a dilute population of interacting swimmers in a quiescent fluid. We introduce the one-point probability distribution function, $\Psi(\rr,\pp,t)$ for swimmers with orientation $\pp$ at position $\rr$ and time $t$.  The dynamics of  the active particles is defined by Eqs~\eqref{eqposition}-\eqref{eqorientation}, with the fluid velocity field, $\uu(\rr,t)$, resulting from the linear superposition  of force dipoles induced by each swimmer,  $\uu(\rr,t)=  \int \!  {\rm d}\pp {\rm d}\rr' \, \Psi ({\rr', \pp},t) \uu^{\rm d}(\rr|\rr', \boldsymbol \sigma')$, where $\boldsymbol \sigma'= \sigma v_{\rm s} \pp$.  Assuming, that swimmers are subject to translational and rotational diffusion, $\Psi(\rr,\pp,t)$ obeys the continuity equation
\begin{equation}
\partial_t \Psi = - \nabla \cdot(\Psi \dot \RR) - \nabla_{\pp}\cdot (\Psi \dot \pp)+D\nabla^2 \Psi+D_r\nabla_\pp^2 \Psi,   
\label{eqpsi}
\end{equation}
where $\dot\RR$ and $\dot\pp$ are defined by Eqs.~\eqref{eqposition}-\eqref{eqorientation},
$D$ and $D_r$ are the translational and the rotational diffusion coefficients respectively, and $\nabla_\pp$ stands for the gradient on the unitary circle. For simplicity, we neglect the translational diffusion. Specifically, anticipating on our results, we assume $D\ll v_{\rm s}^2/D_{R}$, which is true for most biological and artificial micro-size swimmers.  Note that for homogeneous 
suspensions, and due to the symmetry of the dipolar coupling, the sum of all hydrodynamic interactions vanishes: when $\nabla\Psi(\rr,\pp,t)=0$, we have $\int \!   {\rm d}\rr' \,  \uu^{\rm d}(\rr|\rr', \boldsymbol \sigma')=0$, and thus  from Eqs.~\eqref{eqposition}-\eqref{eqorientation} it follows that $\dot\pp=0$, and $\nabla\cdot \dot \RR=0$. The dynamics of an homogeneous  population, from Eq.~\eqref{eqpsi}, reduces thus to the orientational diffusion of an isolated swimmer, and  homogeneous phases  relax toward an isotropic state over a time  $\sim D_R^{-1}$. 

We now investigate the dynamic response of the homogeneous and isotropic phase to spatial fluctuations of the concentration and orientation of the active particles. The phase behavior is described in term of (i) the concentration field, $c(\rr,t)\equiv\int  \! \Psi(\rr,\pp,t){\rm d}\pp$, (ii) the local polarization, $\PP(\rr,t)\equiv\frac{1}{c}\int \! \pp\Psi(\rr,\pp,t){\rm d}\pp$, and (iii) the local nematic-orientation tensor, $\QQ(\rr,t)\equiv\frac{1}{c}\int  \!(\pp\pp-\frac{1}{2}{\bf I})\Psi(\rr,\pp,t){\rm d}\pp$. To establish their equation of motion, we need to add a closure relation to Eq.~\eqref{eqpsi}. As we focus on deviations from isotropic and homogeneous states, 
we expand $\Psi$ linearly in  its three first moments~\cite{Baskaran:2009vu,Leoni:2010bh}
\begin{equation}
\Psi ({\bf x,p},t) = \frac{1}{2\pi} c \left( 1+ 2 p_\alpha P_\alpha + 4 p_\alpha p_\beta Q_{\alpha \beta} \right),
\label{probdistriso}
\end{equation}
where the numerical coefficients are chosen so that  $c$, $\PP$, and $\QQ$ are defined in a self-consistent fashion. Defining $\bar\mu\equiv\frac{1}{2}(\mu_\parallel+\mu_\perp)$, and $\tilde\mu\equiv(\mu_\parallel-\mu_\perp)$, and after some elementary but tedious algebra, the three nonlinear equations of motion are inferred from Eqs.~\eqref{eqpsi}-\eqref{probdistriso} as
\begin{align}
&\partial_t c = - \nabla_\alpha \left[v_{\rm s} c P_\alpha + \bar\mu c u_\alpha+ \tilde\mu c Q_{\alpha \beta} u_\beta\right]\label{cevo},\\
&\partial_t (cP_\alpha) = \frac{\nu}{2} u_\alpha c - \nu c u_\beta  Q_{\beta\alpha}   - D_R c P_\alpha  - \nabla_\beta {\cal I_{\beta\alpha}} \label{polarevo},\\
&\partial_t (c Q_{\alpha \beta}) = \frac{\nu}{2} c u_\gamma ( 2\delta_{\gamma(\alpha} P_{\beta)}-\delta_{\alpha\beta}P_\gamma) 
- 4 D_R c Q_{\alpha\beta}-  \nabla_\gamma {\cal J_{\gamma\alpha\beta}} 
,\label{nemparevo}
 \end{align}
where the (potential) fluid velocity satisfies
\begin{equation}
\partial_\alpha u_\alpha=-\sigma v_{\rm s} \partial_\alpha\left(cP_\alpha\right),
\label{eqvelocitygeneral}
\end{equation}
and where  the expressions for the fluxes $ \boldsymbol {\cal I}$ and $ \boldsymbol{\cal J }$ are given in supplementary information. 

Equations~\eqref{cevo}-\eqref{eqvelocitygeneral}  fully describe the dynamics of the isotropic phase.
We  investigate  their linear stability with respect to  plane-wave excitations of the form $(\delta c, \delta \PP, \delta\QQ)\exp(i\kk\cdot\rr-i\omega t)$, with $\kk=k\hat{\bf x}$.  
   At linear order, we can integrate Eq.~\eqref{eqvelocitygeneral} for the fluid velocity, and recast the equations of motion into a set of two uncoupled linear systems having the form $\partial_t (\delta P_y,\delta Q_{xy})=M_{\rm bend}(\delta P_y,\delta Q_{xy})$ and $\partial_t (\delta c, \delta P_x,\delta Q_{xx})=M_{\rm splay}(\delta c, \delta P_x,\delta Q_{xx})$. The first system couples the transverse-polarization and the bend modes  only. These modes are stable for all $k$, they correspond to damped sound-waves. The associated dispersion relation is deduced from the eigenvalues of $M_{\rm bend}$ as $i\omega=\frac{1}{2}(5D_R\pm i\sqrt{-9D_R^2+(kv_{\rm s}/2)^2})$. In contrast, long-range hydrodynamic interactions between  swimmers can destabilize the concentration ($c$), the longitudinal polarization ($P_x$) and the splay modes ($Q_{xx}$). To convey an intuitive description of this instability we introduce the two governing dimensionless numbers. First, ${\rm Pe}\equiv\nu c_0\sigma v_{\rm s}/(2D_R)$ is a Peclet number comparing  the rotational-diffusion rate $D_R$ to the rate of rotation of a polar swimmer induced by a source dipole of magnitude $\sigma c_0v_{\rm s}$ ($c_0$ being to the average concentration); large-tail swimmers (resp.~large-head swimmers) correspond to ${\rm Pe}>0$ (resp.~${\rm Pe}<0$).
The second dimensionless number, $H\equiv (\bar \mu \sigma c_0v_{\rm s})/v_{\rm s}$, compares  the swimming speed, $v_{\rm s}$, to the advection velocity induced  by a source dipole of magnitude $\sigma c_0v_{\rm s}$. In the long-wave-length limit ($k\to0$), the  eigenfrequencies associated with the stability matrix $M_{\rm splay}$ then take the form 
\begin{eqnarray}
\omega_{c}&=&-i\frac{v_{\rm s}^2}{2D_R}\left(\frac{1-H}{1+{\rm Pe}}\right)k^2,
\label{eigenvaluec1}\\
\omega_{\rm P}&=&-iD_R\left(1+{\rm Pe}\right)+{\cal O}(k^2),
\label{eigenvalueP1}\\
\omega_{\rm Q}&=&-4iD_R+{\cal O}(k^2).
\label{eigenvalueQ1}
\end{eqnarray}

\begin{figure}
\begin{center}
 \includegraphics[width=0.8\columnwidth]{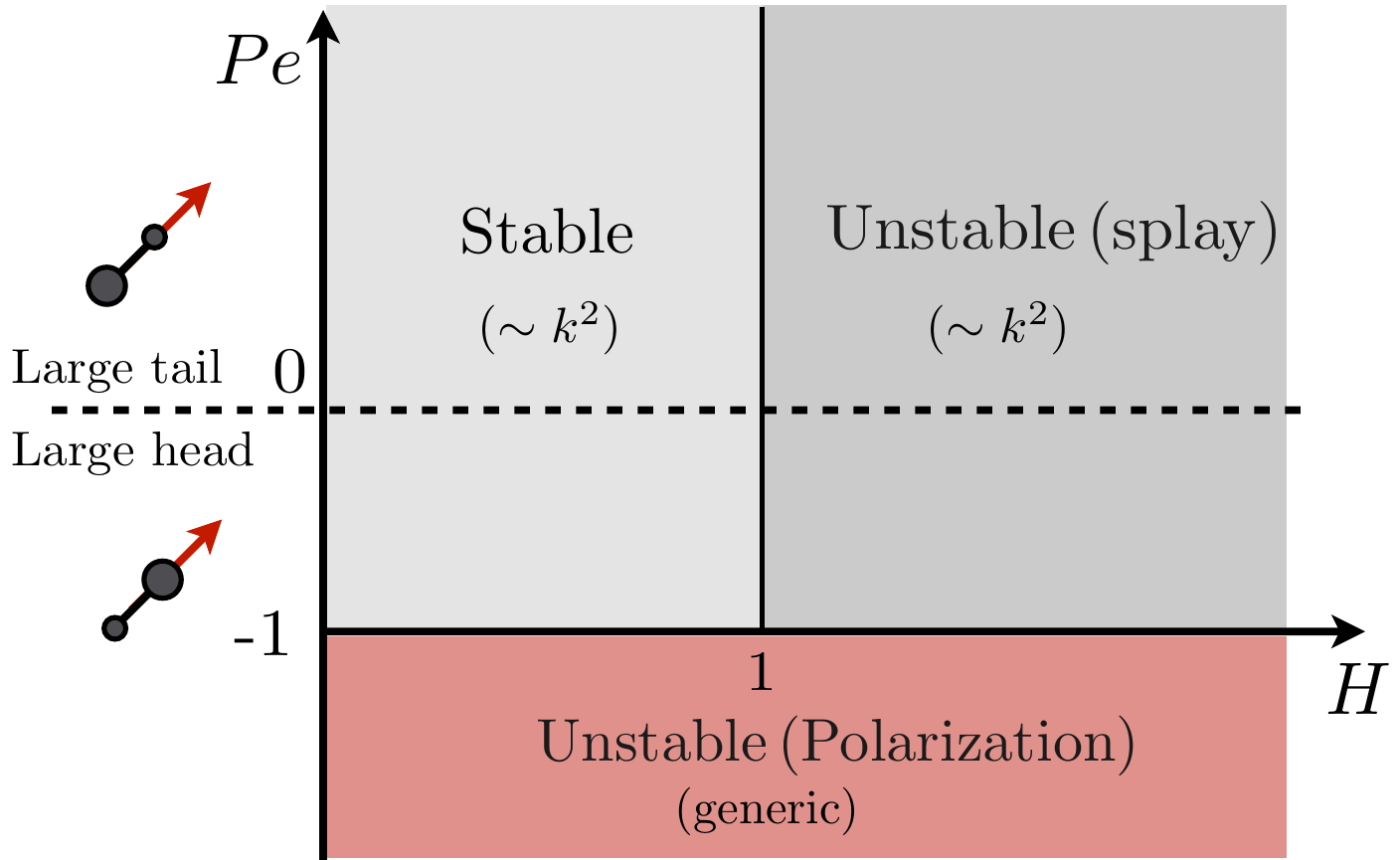} 
\caption{Stability diagram of a nearly isotropic and homogeneous population of polar swimmers; 
${\rm Pe}<0$ (resp.~${\rm Pe}>0$) refers to large-head swimmers (resp.~large-tail swimmers).}
\label{fig2}
\end{center}
\end{figure}
At $0^{\rm th}$ order in $k$, the total number of swimmers being a conserved quantity we have $\omega_c=0$,  and  $M_{\rm splay}$ has only two non-trivial eigenvalues. Whereas  rotational diffusion always stabilizes the nematic orientation ($-i\omega_Q<0$),  hydrodynamic interactions can in fact  destabilize  the isotropic state.  From Eq.~\eqref{eigenvalueP1}, we see that large-head swimmers with ${\rm Pe}<-1$   experience a generic  instability:  fluctuations of the local polarization are amplified when the rotation induced by the hydrodynamic couplings overcome the diffusional relaxation of $P_x$ (see Fig.~\ref{fig2}). 

Several comments are in order. First, although the growth rate of the instability does not dependent on $k$, the total polarization ($k=0$) is  {\em not} unstable. 
As discussed above, the sum of all the hydrodynamic interactions cancels in this limit and no  global  directed flow can  emerge spontaneously from an isotropic suspension. The instability shows however that groups of particles swimming coherently along the same direction  form at all scales. Second, the generic nature of the instability is specific to   the dipolar symmetry of the hydrodynamic interactions, and  the polar shape of the particles, and can be intuitively rationalized as follows.  From Eq.~\eqref{eqvelocitygeneral} we see that any finite wave-length perturbation of $P_x$ along $x$ results in a fluid flow in the opposite direction, with  amplitude $\sim \sigma c_0v_{\rm s}\delta P_x$. Polar swimmers  align with, or against, the local  flow direction depending on their polarity. Large-head swimmers align along $-\uu$, thereby increasing the initial perturbation of $\PP$ and destabilizing the isotropic state. Conversely, large-tail swimmers align in the opposite direction and the local polarization  relaxes to zero. As the reorientation rate of the  swimmers is set by the magnitude of the velocity only (and not by the local strain-rate tensor), the growth  (or relaxation) rate of the polarization  is independent of the wave vector.

This novel generic instability is qualitatively different from the one observed in unbounded suspensions of pushers which, in contrast, is suppressed by confinement~\cite{Marchetti:2012ws}. They differ in both the physical mechanisms at work and the structure of the unstable modes (bend versus splay modes). The only similarity is that in both systems the generic  instability is a genuine collective effect due to  the long-range nature of hydrodynamic interactions. 

To investigate the stability of the active film when ${\rm Pe}>-1$, we need to consider the eigenfrequencies , and the eigenmodes of $M_{\rm splay}$ up to  $O(k^2)$.  From Eq.~\eqref{eigenvaluec1} we see that the combination of self-propulsion and rotational diffusion yields an effective diffusive dynamics of the suspension scaling as $\omega_c\sim (v_{\rm s}^2/D_R)k^2$, as could have been anticipated from the single swimmer problem~\cite{Howse:2007ed}. However, hydrodynamic interactions result in a renormalization of this single-swimmer effect. These interactions control both the magnitude and the sign of the effective  translational diffusion. In the regions (${\rm Pe}>-1$, $H>1$) and (${\rm Pe}<-1$, $H<1$),  
the effective diffusivity is negative and  thus slowly destabilizes the isotropic phase (Fig.~\ref{fig2}). The associated eigenmodes are now complex superpositions of $c$, $P_x$, and $Q_{xx}$, and thus  clusters of aligned particles form and propel in a coherent fashion (swarms), from a homogeneous film. Notably,  both large-head  ($-1 < \rm Pe < 0$) and large-tail ($\rm Pe > 0$) swimmers are prone to this second splay-destabilization mechanism. In the other regions of Fig.~\ref{fig2}, the effective diffusivity is positive and  concentration fluctuations are stable.

In summary we   revisited the theoretical description of confined populations of micro-swimmers. We  showed that active particles interact hydrodynamically in generic manner, 
which is independent of the microscopic details of their propulsion mechanism and that, depending on their polarity:  they may reorient in flows instead of solely flow gradients.  Focusing on polar swimmers, we then constructed a large scale hydrodynamic theory from a minimal microscopic model (dummbells). Our analysis showed that the macroscopic orientational dynamics is very different from the modified Leslie-Eriksen model of active liquid crystals due to a difference in the symmetry of the microscopic coupling between confined polar particles and the fluid flow. It results in a novel phase behavior  for active films and, in particular,  spontaneous  large-scale directed motion and swarming can emerge out of isotropic populations of confined swimmers. 

This work was funded in part by the NSF (grant 0746285 to E.L.), Paris Emergence research program (D. B.), and C’Nano Idf (D. B.).
We thank  Aparna Baskaran, Olivier Dauchot  and David Saintillan fro valuable discussions.

\begin{widetext}
{\bf Supplementary informations:} Using  Ricci-calculus notations, the expression of the fluxes $ \boldsymbol {\cal I}$ and $ \boldsymbol{\cal J }$ are:
  \begin{align}
{\cal I_{\beta\alpha}}&=     \frac{1}{4} \left[ \tilde \mu c P_\gamma u_\gamma \delta_{\alpha \beta}  
+(4\bar \mu-\tilde \mu) c P_\alpha u_\beta 
+\tilde \mu c P_\beta u_\alpha +
 4\nu cv_{\rm s} (Q_{\alpha \beta} + \frac{\delta_{\alpha \beta}}{2} ) \right],\\
{\cal J_{\gamma\alpha\beta}} &=  \frac{v_{\rm s}  c}{2} [  \delta_{\gamma(\alpha} P_{\beta)}-\frac{\delta_{\alpha\beta}}{2}P_\gamma ]
-  \frac{\tilde\mu c}{4}[6 u_\gamma \delta_{\alpha\beta} - u_{(\alpha} \delta_{\beta)\gamma}]%
+2\bar\mu u_\gamma c Q_{\alpha\beta}\nonumber \\
&+\frac{1}{6}\tilde\mu c\left[
-4u_\gamma Q_{\alpha\beta}-5u_\delta  Q_{\delta\gamma} \delta_{\alpha\beta}
\right.+\left .
2 u_\delta Q_{\delta(\alpha} \delta_{\beta)\gamma}+2u_{(\alpha}  Q_{\beta)\gamma} 
\right],
\end{align}
\end{widetext}
\end{document}